\documentclass[12pt]{article}

\usepackage{amssymb}

\usepackage{graphicx}
\usepackage{epstopdf}

\usepackage{times}

\title{Light-Activated Self-Propelled Colloids }

\author{
J. Palacci$^{1*}$, S. Sacanna$^{2}$, S.-H. Kim$^3$, G.-R. Yi$^{3}$,\\ D.J. Pine$^{1}$ and P.M. Chaikin$^{1}$ \\
\normalsize{$^{1}$Department of Physics, New York University, USA}\\ 
\normalsize{$^{2}$ Department of Chemistry, New York University, USA}\\ 
\normalsize{$^{3}$ School of Chemical Engineering, Sungkyunkwan University, Republic of Korea} \\
\normalsize{$^\ast$To whom correspondence should be addressed; Jeremie Palacci: jp153@nyu.edu.}
}

\begin{document}

\baselineskip24pt


\maketitle 

{\bf Abstract \\}
Light-activated self-propelled colloids are synthesized and their active motion is studied using optical microscopy. We propose a versatile route using different photoactive materials, and demonstrate a multi-wavelength activation and propulsion. Thanks to the photo-electrochemical properties of two semi-conductor materials ($\alpha$-Fe$_2$O$_3$ and TiO$_2$), a light with an energy higher than the bandgap triggers the reaction of decomposition of hydrogen peroxide and produces a chemical cloud around the particle. It  induces a phoretic attraction with neighboring colloids as well as an osmotic self-propulsion of the particle on the substrate. We  use these mechanisms to form colloidal cargos as well as self-propelled particles where the light-activated component is embedded into a dielectric sphere. The particles are self-propelled along a direction otherwise randomized by thermal fluctuations, and exhibit a persistent random walk. For sufficient surface density, the particles spontaneously form  "living crystals" which are mobile, break appart and reform. Steering the particle with an external magnetic field, we show that the formation of the dense phase results from the collisions heads-on of the particles.  This effect is intrinsically non-equilibrium and  a  novel principle of organization for systems without detailed balance. 
Engineering families of particles self-propelled by different wavelength demonstrates a good understanding of both the physics and the chemistry behind the system and points to a general route for designing new families of self-propelled particles.
\\


\section{Introduction}
Shrinking people down to the micron is a classical science-fiction premise in which the agents could 
manipulate tiny objects  as in a microscopic factory. Ultimately, they may be injected in the body and repair disfunctional organs or carry drugs to the appropriate cells. 
Beyond the limitless imagination of writers, this points towards a challenging question for the scientific community: how can we design populations of artificial 
micro-agents capable of moving autonomously in a 
controlled fashion while performing complex tasks?
 Recently, this question has fueled a great effort towards the fabrication and the development of the first generation of synthetic micro and nano robots \cite{Wang:2013cc, Sengupta:2012ie, Wang:2012kd}. \\
In living matter, the energy is extracted from the hydrolysis of ATP into ADP, which constitutes the "quantum" of intracellular energy transfer. Similarly, artificial systems need to harvest the free energy from the environment and convert it into mechanical work. Numerous experimental realizations of such systems have been performed in recent years, many of them, taking advantage of  phoretic mechanisms \cite{anderson, CordovaFigueroa:2008db, Paxton:2004ea, Howse:2007ed, Pavlick:2011ba, Jiang:2010el, Palacci:2010hk, Baraban:2013ef}, which interfacial origin provides a driving force robust to downsizing.
Collections of active micro-particles thereafter constitute a controlled realization of active matter,  in which self-driven units convert an energy source into useful motion and work, and provide a formidable playground for the study of phenomena in internally driven systems.  Active matter  exhibits a wealth of non-equilibrium effects observed in nature as well as synthetic systems: pattern formation \cite{Budrene:1995ew,Budrene:1991gq}, enhanced mixing \cite{PhysRevLett.84.3017, Leptos:2009kd}, or sensing and interaction with the environment. For example,  E. Coli bacteria were shown to concentrate due to the presence of microfluidic funnels \cite{PeterGalajda12012007}, rotate microscopic gears \cite{DiLeonardo:2010kk,Sokolov:2010kt} or self-propelled nanorods are captured by passive spheres, stressing the importance of activity-driven interactions \cite{Takagi:2014hf} . From a fundamental standpoint, they allow for the development of a theoretical framework for non-equilibrium statistical mechanics \cite{Ramaswamy:2010bf}.\\

In  this paper, we focus on recent developments on light-activated  self-propelled colloids using  the photocatalytic decomposition of hydrogen peroxide as a source of free energy. 
We present various realizations of self-propelled colloids using different  photocatalytic  materials, titanium oxide and hematite,  which can be activated by the adequate wavelength.  The activation induces  interfacial flows leading to an osmotic self-propulsion of the particles and a phoretic attraction between them.  We harness these mechanisms to use them as colloidal dockers to target and transport passive objects at the microscale. Finally, we discuss the collective properties of a dense  suspension of these spherical active particles. They form "living crystals" which form, break apart, heal and reform. We reproduce these crystals using simple numerical simulations of self-propelled hard disks coupled by a phoretic attraction. We show that the collisions are required to account for the observed effect, stressing the role of self-trapping as an organization principle in non-equilibrium systems. 

 \section{Photo-activated colloids}
 \subsection{Photocatalytic materials}
 The particles considered in this paper are micron-size particles containing a photo-active materials, immersed in a fuel solution containing hydrogen peroxide. Since hydrogen peroxide is  metastable at ambient temperature, the reaction of decomposition, 2H$_2$O$_2=2$H$_2$O + O$_2$, is thermodynamically favorable but kinetically limited. The energy barrier is overcome by providing external energy such as  UV light or heat  or by the presence of a catalyst lowering the energy barrier. 
Though the mechanism for the photocatalytic decomposition of hydrogen peroxide is not well understood, we infer that the effect originates in the reaction of the hydrogen peroxide in solution with the electron-hole pairs  generated by the absorption of ultra-band gap energy photons in  the semi-conductor (SC) \cite{ZHANG:1993vr}.  The reaction produces radicals that cause the rapid  degradation and bleaching of  fluorescent dyes present in solution. A similar effect is observed  by  Soler {\it et al}, who used Fenton reactions over Fe-Pt nanorods to degrade Rhodamine 6G in a solution of hydrogen peroxide  {\it et al.} \cite{Soler:2013kg}. \\
 
In this paper, we present particles with two different SC as photo-active materials.  We use an iron oxide, ($\alpha$-Fe$_2$O$_3$), named hematite  [Fig.1A], with a band gap of about 2.2eV, corresponding to a visible wavelength   $\lambda_{Fe_2O_3}\sim 560$nm \cite{Xia:2013gt} and previously studied in \cite{Palacci:2013eu,Palacci:2013tu}.
 We moreover present a new colloidal particle taking advantage of an alternate photo-active material:  titania  (TiO$_2$) in the anatase phase [Fig.1B] , with a bandgap of 3.1eV, corresponding to the wavelength   $\lambda_{TiO_2}=400$nm \cite{Hegazy:2012fx}. \\ 
 The  particles  are conveniently observed with an inverted optical microscope equipped with high magnification objectives, typically 60x or 100x and a conventional bright-field diascopic illumination. The microscope is equipped with   filters to combine the bright field illumination with various wavelength bands from an episcopic fluorescent lamp (ultra high pressure 130W mercury lamp, Nikon Intensilight), filtered and focused on the sample through the observation objective. We use interchangeable bandpass filters to select  the windows of wavelength  of the excitations $\lambda_E$. In the experiment, we use bandpass filters with an excitation  E$_1$, in the blue   ($\lambda_{E_1}\in[430\mathrm{nm}-490\mathrm{nm}]$, {Semrock}, FF01-460/60-25), or E$_2$, UVA-violet, ($\lambda_{E_2}\in [370\mathrm{nm}-410\mathrm{nm}]$, { Semrock}, LF405/LP-B).   We can manually swap the filters during the experiment.  A mechanical shutter on the lamp allows to turn on an off the excitation light providing a  wireless and reversible activation of the system [Fig.1C].
  \begin{figure}[htbp]
\centering\includegraphics[width=4in]{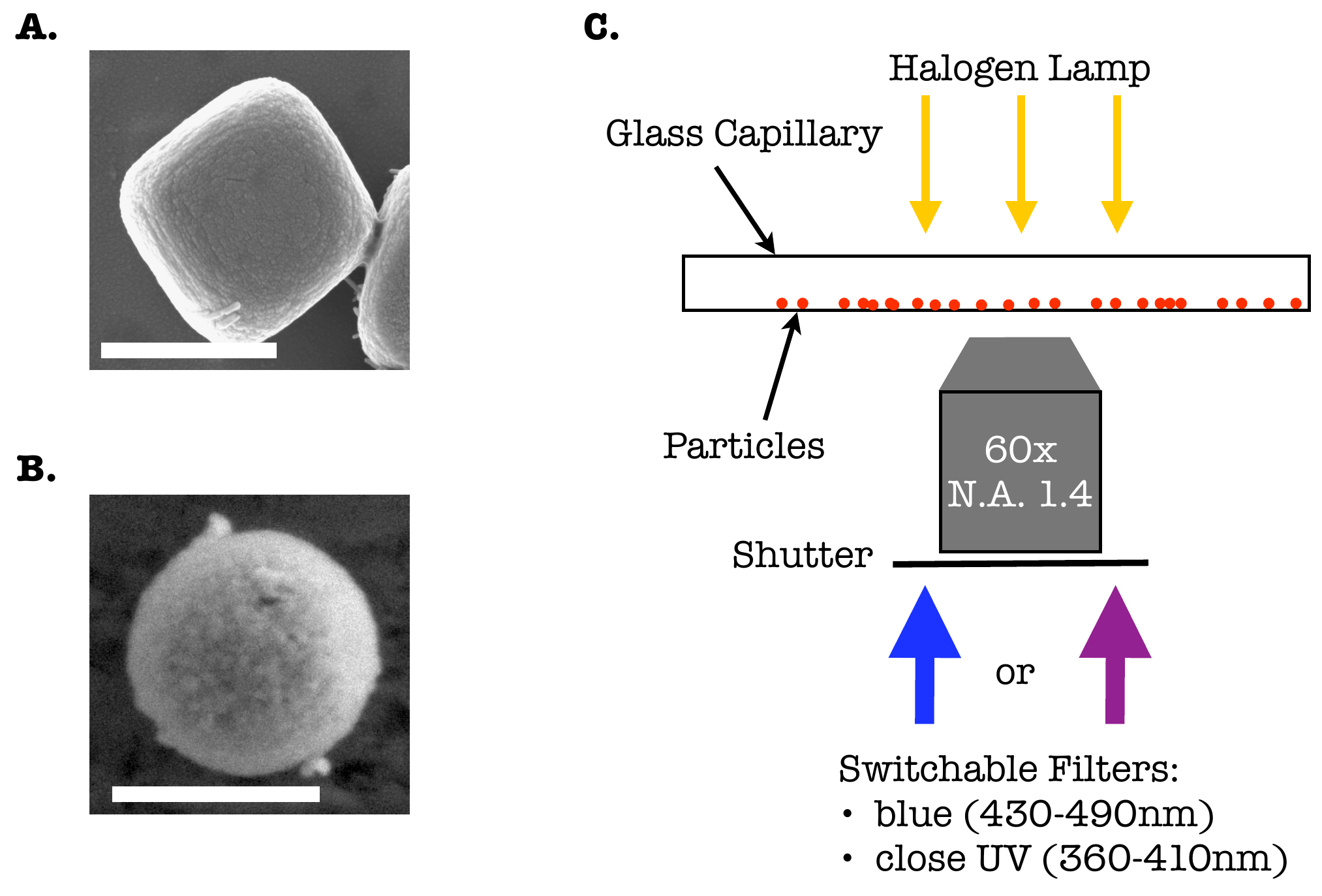}
\caption{ {\bf A.} Scanning Electronic Microscopy (SEM) picture of a photoactive hematite cube. Scale bar is 500nm. {\bf B. }SEM picture of a photoactive titania particle. Scale bar is 500nm. {\bf C.} Experimental setup. A capillary is filled with the solution of particles and observed in an inverted microscope. The excitation light comes from a filtered UV light source through a large magnification objective. The excitation wavelength can be changed  manually with bandpass filters. Here we use  blue (430-490nm) light or UVA-violet (370-410nm). A mechanical shutter allows for an external and wireless actuation of the system.} 
\label{fig1}
\end{figure}
\vspace*{-10pt}

\subsection{Chemical gradients and diffusio-phoresis}
\subsubsection{Chemical Gradient}
The photocatalytic material is immersed in a solution of  hydrogen peroxide fuel H$_2$O$_2$. Under activation by light, it decomposes the hydrogen peroxide fuel and the concentration profile  of hydrogen peroxide, $[H_2O_2]_{(r, t)}$,  is given by the solution of the diffusion-reaction equation:
\begin{equation}
\partial_t [H_2O_2]_{(r, t)}= D^{\star}\Delta [H_2O_2]_{(r, t) } - \alpha  [H_2O_2]_{(0, t) }  
\label{Eq:diff}
\end{equation}
where $t$ is the time,  $D^{\star}$ is the self-diffusion constant of hydrogen peroxide and $\alpha$, the reaction rate of decomposition of hydrogen peroxide on the considered photocatalytic surface, at position $r=0$. \\
At steady state, Eq \ref{Eq:diff} simplifies to the Laplace equation, $D^{\star}\Delta [H_2O_2]_{(r)} =\alpha  [H_2O_2]_{0} $. The solution of this equation in 3D, for an infinity large reservoir of hydrogen peroxide, in the diffusion-limited regime is:
\begin{equation}
[H_2O_2]_{(r)}= [H_2O_2]_{\infty} (1-b/r)
\label{Eq:diff2}
\end{equation}
 where $[H_2O_2]_{\infty}$ is the bulk concentration of hydrogen peroxide and $b$ the half size of the catalytic site. The photocatalytic material acts as a sink for the hydrogen peroxide and a source of oxygen $O_2$, which dissolves in the solution [Fig2.a, b]. Following the stoechiometry of the decomposition reaction,  Eq. \ref{Eq:diff2} give the concentration of oxygen in the solution: $[O_2]_{(r)}= \frac{ [H_2O_2]_{\infty}}{2} \times b/r$.

 \subsubsection{Effect of the chemical gradient: diffusio-phoresis}
In order to study the effect of the chemical gradient, a hematite particle is attached to the bottom surface of a glass capillary and immersed in a solution of hydrogen peroxide containing the fuel (hydrogen peroxide) and TMAH, at pH$\sim 8.5$. The hematite particle is immobilized  and conveniently observed under an optical light microscope.  \\
 In the absence of activation by light, the colloidal tracers diffuse in the solution, at equilibrium with the solvent. Under light activation (blue or UVA-violet for the hematite, or UVA- violet for the titania particles), the colloids in the solution are attracted towards the photoactive materials, for all the material we tested (various polymeric colloids or silica). The attraction comes from every direction,  thus discarding the possibility  of a flow  which would exhibit recirculation by incompressibility of the fluid [Fig.2B]. Additional experiments with a hematite cube sedimenting through a solution of colloids also shows an isotropic attraction, ultimately leading to the formation of raspberry-like particle, with the photo-active material at the core. \\
 The colloidal particles are migrating in response to the chemical gradients, oxygen (O$_2$) or hydrogen peroxide (H$_2$O$_2$). The motion of colloids induced by a solute gradient is called   {\it diffusiophoresis}.  It is an interfacial physical mechanism which belongs to the more general class of surface-driven phoretic phenomena \cite{anderson, CordovaFigueroa:2008db}.   It results from an unbalanced osmotic pressure occurring within the diffuse layer in the close vicinity of a solid surface (typically of the order of a few nanometers), which thereby plays the role of the semi-permeable membrane in the classical osmosis. This induces an interfacial flow along the surface leading to the motion of the particle in the surrounding medium \cite{ISI:000259445400018, Palacci:2012gk}. Diffusiophoresis is therefore material and pH dependent, since it originates from the interaction between the solute and the solid surface. \\
We gather the results of the migration of $1.5\mu$m colloids made out of 3-methacryloxypropyl trimethoxysilane (TPM), a polymer material,  to a hematite photoactive cube [Fig.2C]. We record the displacement of an ensemble of particles (typically 10) towards the hematite particle under light activation by blue light, excitation $E_1$. The different trajectories are averaged out in order to extract the diffusio-phoretic drift R(t) from the random Brownian noise [Fig.2C, black filled  symbols). We differentiate the experimental curve  to compute the   the phoretic velocity, $V_p(r)$, at a distance $r$ from the center of the cube to the center of the sphere. We measure a good agreement between the experimental measurement $V(r)$ and the diffusiophoretic velocity in a solute gradient $C(r)$, for which $V_{DP}\propto \nabla C(r)\propto 1/r^2$ [Fig.2D, red dashed line]. \\
Reproducing the experiment with spheres of different materials (TPM, polystyrene (PS), or silica), we observe a dependence of the attraction with respect to the material [Fig.2E]. However, for the different materials tested, the attraction strength is  of the same order of magnitude  and we measure diffusiophoretic mobilities within a factor of 2.  \\
In our experiments, we did not find any material which did not present attraction towards the hematite at the considered pH$\sim 8.5$: TPM, PS, silica, PDMS, quartz...
However, the diffusiophoretic migration is reversed for TPM and PS if we lower the pH to 6.5 by suppressing the TMAH in solution: the colloidal particles are repelled away from the photoactive material. \\

  \begin{figure}[htbp]
\centering\includegraphics[width=5in]{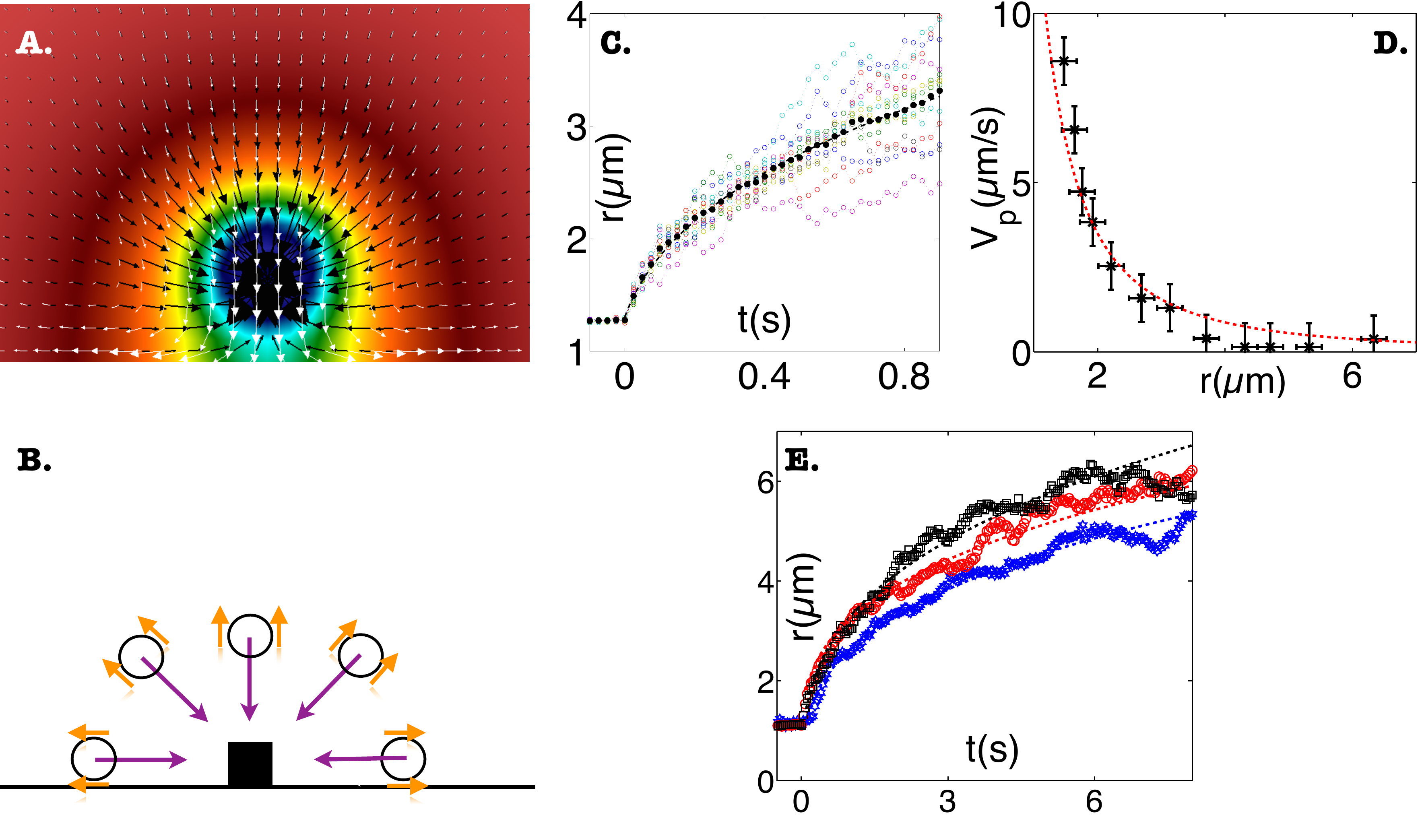}
\caption{{\bf A.} 3D simulations of the experiment of a photoactive particle on a surface. The colormap represents  the concentration of fuel, here H$_2$O$_2$ (linear scale). It follows a $\propto 1/r$ decay.  The white arrows represent the osmotic pumping flow along the substrate induced by the particle. The black arrows is the velocity of a phoretic particle in the solution, thus a superimposition of the phoretic migration of the particle  and the advection by the osmotic flow. The phoretic migration dominates the osmotic contribution for particles not strictly restricted to the very close vicinity of the wall. The length of the arrows represents the intensity of the flow. Simulations realized by the group of A. Donev (Courant Institute, NYU). {\bf B.} Sketch of the experiment. The particles in solution sense the chemical gradients. They exhibit an interfacial flow (orange arrows) resulting in a diffusiophoretic migration (purple arrows) towards the photoactive material. The attraction is isotropic, particles come from every direction, discarding the possibility of an hydrodynamic flow of an incompressible fluid. {\bf C.} Timelapse displacement of different 1$.5 \mu m$ TPM colloids in a concentration gradients (colored symbols). The trajectories are averaged out to suppress the thermal random component of the displacement (black full symbol) and extract the phoretic drift. {\bf D.} The migration velocity $V_p (r)$ (black symbols) is extracted by differentiation of the average displacement in C. For a diffusio-phoretic migration, we expect a migration velocity proportional to the gradient of the concentration, hence $\propto 1/r^2$ (red dashed line), showing a good agreement with the experimental data. {\bf E.} Phoretic migration for tracers of $\sim 1.5\mu m$ size made of different materials: silica (blue symbols), TPM (red symbols), PS (black symbols). The experimental data are fit by $ r(t)=  A t^{1/3}$, prescribed by a diffusiophoretic mechanism. The diffusiophoretic mobility depends on the material and ranges within a factor of two for the considered materials. }
\label{fig2}
\end{figure}
\vspace*{-10pt}
\subsubsection{Self-propelled osmotic surfers}
We now consider the situation of a photo-active material (hematite or titania) dispersed in a solution of fuel with hydrogen peroxide and pH$\sim 8.5$. The particles are heavy and reside near the bottom surface. In the absence of any photo-excitation,  those are Brownian particles, diffusing near the bottom surface. If we shine the appropriate excitation light, it triggers the chemical reaction of decomposition and the particle establishes a chemical cloud $C(r)$ in its surrounding, following \ref{Eq:diff2}. In the same way a free colloidal particle migrates in a gradient by {\it phoresis}, a fixed surface of the same material induces an {\it osmotic flow} in opposite direction when exposed to a gradient \cite{anderson}. The glass substrate of the capillary is exposed to the concentration gradient generated by the active particle, and induces  an osmotic pumping flow, attracting the particles to the surface [Fig.2A, white arrows]. The surrounding chemical cloud is in principle symmetric and the particle should sit there. However, we observe experimentally that a significant portion of the particles starts propelling on the substrate [movie 1] .\\ 
We envision two different {\it scenarii} to account for this effect:  (i) the particles are not perfectly symmetric, one side is more chemically active than the other, thus breaking the symmetry. This could originate in "imperfections" from the synthesis, or intrinsic anisotropic properties of the particles. For example, it is known that the particles surface has an intrinsic chemical anisotropy due to the fine grained structure of the hematite \cite{SHINDO:1993uj}. 
Unfortunately, the size of the particles --typically 600nm--  challenges the optical resolution of the microscope and  it is  difficult to resolve experimentally the facets of the cube making this hypothesis difficult to test in a direct observation.  However, we observe that an increase of the roughness of the particles by a strong acid treatment (1M hypochloric acid)  favors the propulsion of particles. This points towards the importance of  roughness and chemical anisotropy in the system. \\
An alternate scenario (ii) is a spontaneous symmetry breaking mechanism. In a nutshell, the particle sits in a symmetric gradient, until a fluctuation pushes it on one direction, the chemical gradient is steepened on this side and smoothened on the other side, breaking the symmetry and the particle  follows the direction of the fluctuation. Such spontaneous autophoretic propulsion of isotropic phoretic particles has been discussed numerically by Michelin {\it et al} \cite{Michelin:2013gv}. The authors show there that isotropic particles can exhibit spontaneous self-diffusiophoretic propulsion and spontaneous  symmetry-breaking  as a result of an instability driven by the Peclet number $Pe$ of the system (the Peclet number compares the role of advection and diffusion in the transport of the solute: $Pe=R\times V_{prop}/D$). Our experiments are typically at very low Peclet:  the diffusion is too fast to allow any significant deformation of the chemical cloud due to the propulsion of the particle.  As a consequence, we do not expect this effect to be predominant in the experiment.\\

As a  remark, one could argue that the experiments presented on [Fig.2B,C], with a hematite cube attached to a glass substrate, should exhibit the superimposition  of the osmotic flow  along the substrate with the phoretic attraction, while our earlier interpretation only discusses the role of phoresis [Fig.2D]. Additional numerical simulations of these experiments were performed in the group  of A. Donev (Courant Institute, NYU) to test the effect of the presence of an osmotic flow on the measurement.  An immersed-boundary method \cite{Bhalla:2013di} is used to solve the concentration distribution in the solution [Fig.2A, colormap].  The steady Stokes equation is solved numerically with a slip boundary condition on the bottom wall proportional to the gradient of concentration to obtain the osmotic flow velocity \cite{Griffith:2007do} [Fig.2A, white arrows]. It exhibits a fast decay from the wall and the simulation shows that, at the exception of a very narrow layer near the wall, the phoretic migration totally overcomes the opposite osmotic flow [Fig.2A, black arrows]. This legitimates our measurement in the previous section and stresses the difficulty to measure experimentally the osmotic flow: small colloids will not be restrained by gravity to the close vicinity of the wall while large colloids migration will be dominated by phoresis. One solution is to use phoretically neutral particles. One way to obtain them is to  use "hairy" colloids for which the viscous drag in a dense interfacial polymer forest zeroes the interfacial flow and the phoretic  mobility. Further work is conducted along this line to obtain the colloids and measure the osmotic flow.

\subsubsection{Multiwavelength activation}
The particles are activated using a commercial  fluorescent lamp equipped with bandpass filters. Using an excitation wavelength below the bandgap has no effect on the dynamics of the particles: TiO$_2$ particles, for example,  remain at equilibrium and exhibit a thermal diffusive motion when exposed to the blue light  of excitation spectrum E$_1$: the energy of the blue photons is below the energy bandgap of the material. However, they start propelling once exposed to the UV-violet light E$_2$ in the presence of hydrogen peroxide [movie 2]. According to previous reports  \cite{Hong:2010ita}, titania particles self-propel in pure water under strong UV-light  due to the reaction of water-splitting but we do not observe this effect, in our experimental conditions probably because of the low light intensity and longer wavelengths.\\  Hematite composites are activated by both the excitation E$_1$ and E$_2$. Both violet (or UVA) and blue light are above the energy bandgap of the hematite and the particles exhibit self-propulsion [movie 2].\\
By turning off the excitation light, the system goes back to equilibrium in a few tens of milliseconds and the particles recover a thermal Brownian motion. 
The ability to turn on and off the activity and self-propulsion is an asset of our experimental system, and allows us to discriminate non-equilibrium effects from colloidal aggregation and instability due to the chemical reaction. The multi-wavelength activation is unique and allows to design separate families of active particles.

\subsection{Composite colloids}
\subsubsection{Colloidal Cargos}
In this section, we harness both the phoretic attraction and the osmotic self-propulsion to form composite particles and colloidal cargos. Here we use a mixed system of colloidal particles, typically PS or TPM polymer particles of $1\mu m$ radius, and light-activated colloids, hematite or titania. In the absence of any excitation light, the system is at equilibrium and the particle diffuses. Under activation by light, the light activated particles generate a chemical cloud which can be sensed by a "passive" colloid, then migrating  phoretically towards the photoactive particle. Once docked, the composite system propels as a whole, the photoactive part heading [movie 1 and 3].  The magnetic properties of hematite can furthermore be used to steer and direct remotely the colloidal cargo, as discussed in \cite{Palacci:2013tu} [Fig.3A and movie 3]. One limit of this composite system is that a single passive colloid can couple with many active patches or alternatively one active patch with many passive colloids leading to a complicated and uncontrolled mix of  self-assembled particles, some of them exhibiting self-propulsion, some of them inactivated by symmetry, e.g. one active patch surrounded by four spheres, as theoretically discussed in a recent paper (\cite{Soto:2014fk} [movie 1]. \\
In order to overcome this limitation, we design component particles in which the active patch (hematite or titania) is embedded into a dielectric  shell, protruding outside [see Fig.3 B,C]. These particles are synthesized in bulk, in a controlled manner.
  \begin{figure}[htbp]
\centering\includegraphics[width=5in]{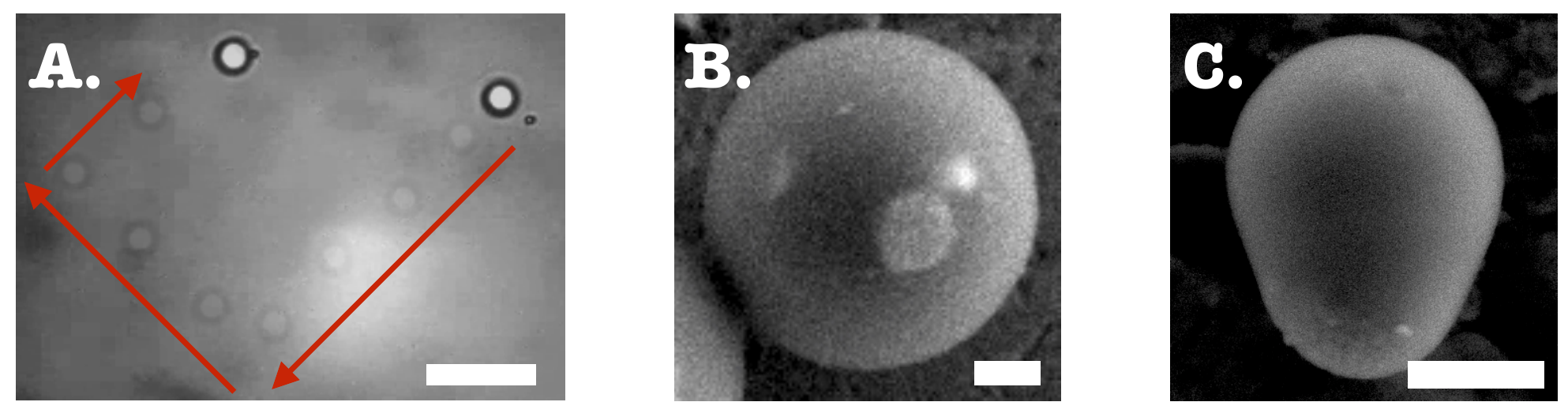}
\caption{{\bf A. Colloidal Cargos:} Timelapse of the trajectory of a colloidal cargo, light spheres are positions every $1s$, showing the initial and final positions. A hematite particle is activated by (blue) light, phoretically attracts a $5\mu m$ TPM sphere, docks it, forming a self-propelling colloidal cargo. The system self-propels as a whole, directed  with an external uniform magnetic field (red arrows). Scale bar is 20$\mu m$. {\bf B.} SEM picture of photoactive hematite particle embedded in a polymer spherical shell. {\bf C.} SEM picture of photoactive titania particle embedded in a polymer  TPM shell.}  Scale bar is 500nm.
\label{fig3}
\end{figure}

\vspace*{-10pt}

\subsection{Colloidal surfers}
Composite colloids, or colloidal surfers, are prepared. They  consist of photoactive materials (hematite or titania) partially protruding outside a shell of TPM. 
The synthesis of these composites is described in details in  the Materials and Methods section. In a nutshell,  we trap solid inorganic colloids (hematite or titania) at the interface of an oil-in-water emulsion droplets of TPM \cite{LEVINE:1989wl}.\\
The hematite iron oxide can be synthesized and obtained in various sizes and shapes (cubes, peanuts or ellipsoids). Seeded-growth of TPM on these particles result in  various particles  as presented in a recent paper by Sacanna {\it et al} \cite{Sacanna:2013hw}.  In this paper,  for an observation with optical microscopy, relatively large  (above 500nm) hematite cubes or ellipsoidal particles are encapsulated as shown in  [Fig.4A-B].\\
Hematite is a canted anti-ferromagnetic iron oxide with a permanent magnetic moment $\vec{\mu}$, scaling as the volume of the hematite component of the particle. Composite particles with  large hematite components interact magnetically and self-assemble in equilibrium dipolar structures  \cite{Sacanna:2012ku}. In the experiment, we limit the size of the hematite component to a typical volume of $ 0.2\mu m^3$ to avoid  magnetic interactions between the particles. The magnetic moment is however large enough to interact with a weak and uniform external magnetic field $B_0\sim$1 mT, tilting the orientation of the cube and allowing to steer the particles externally, as discussed in the previous section.
\\
The composite particles are mixed with the regular fuel solution, in a basic solution (pH$\sim8.5$) containing hydrogen peroxide [0.1 to 3\% weight/weight (w/w)] and 5 mM tetramethylammonium hydroxide (TMAH) in deionized water. The colloids sediment under gravity and reside near the surface of  a clean glass capillary. Under normal bright field illumination, the particles are at equilibrium with their solvent and exhibit a thermal  Brownian diffusion. The rotational diffusion of the particle is visible thanks to the optical contrast provided by the photoactive part. Under light-activation above the bandgap of the material,  the composite particle generates a chemical gradient, inducing an osmotic flow along the substrate, which forces the particle to rotate and the active part to face the substrate.  The particles do not self-propel in bulk and only propel at the substrate. It is usually at the bottom substrate of the cell but it is not restricted  to it by gravity. Particles can climb up the lateral walls of the capillary, the photocatalytic material facing the substrate, as well as propelling upside down on the top surface [movie 4]! Turning off the light, the propulsion stops and the particles sediment towards the bottom surface. \\
We can activate the titania and hematite particles simultaneously using UVA-violet light or only the hematite composite particles shining the blue light on the sample [movie 5]. 
The possibility to use different wavelength to  activate independently different populations of active particles is unique.
Engineering families of particles activated by different wavelength, using different photo-active materials,  demonstrates a good understanding of the system and points to a general route for designing new families of self-propelled particles.\\
The mode of self-propulsion is unusual. Defining a North-South axis along the asymmetry of the particle, the direction of the propulsion is along the equatorial direction. Janus particles  generate a gradient along the asymmetry "pole" axis and subsequently self-propel along this axis \cite{janus_propulsion_ajdari, Golestanian:2007hu}. Here the particles generate the gradient thanks to their photoactive component and harvest the free energy from their environment but the actual engine for the propulsion is localized on the substrate through the osmotic flow.  Grafting a polymer brush to the wall suppresses the osmotic flow at the substrate and therefore the self-propulsion. Alternatively, we observe a $\sim 35\%$ increase of the propulsion velocity of the particles in experiments where the glass capillary are plasma cleaned and {\it immediately} used. Plasma treatments are known to  enhance the charge of glass substrates thus increasing the interfacial transport along it.  After 2 hours, the velocity returns to its original value.  \\
This peculiar mode of propulsion makes the particles very sensitive to the detailed properties and alterations of the substrate. They follow shallow cracks or atomic steps on a cleaved mica, as well as a nano texture imprinted in a PDMS substrate, which would be otherwise ignored by thermal diffusion.
 \begin{figure}[htbp]
\centering\includegraphics[width=5in]{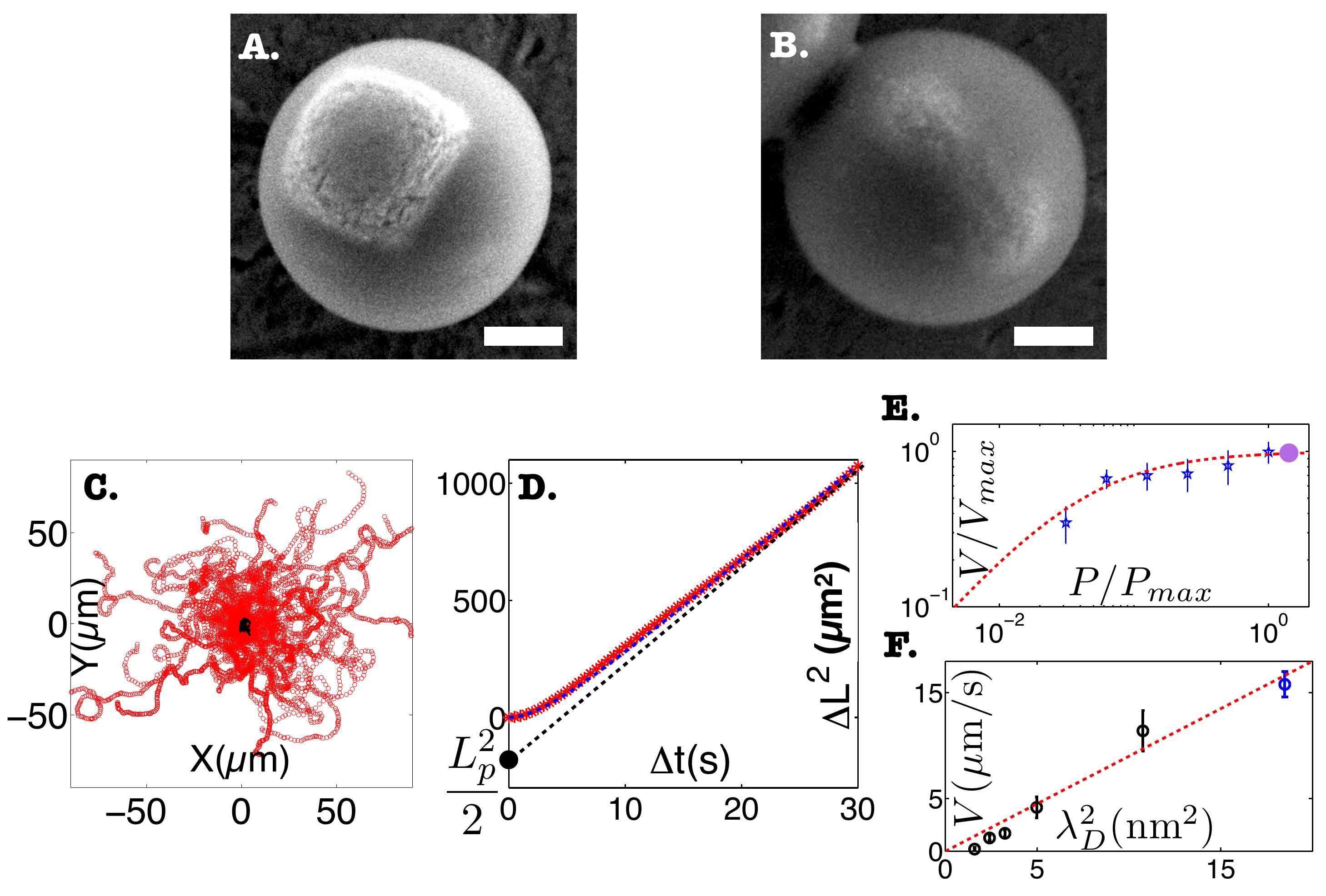}
\caption{{\bf A-B} A diverse family of colloidal surfers. {\bf A.} SEM  picture of large hematite cube embedded in a spherical TPM shell. {\bf B.} SEM  picture of ellipsoid hematite embedded in a TPM shell. {\bf C.} Superimposition of many trajectories of the colloidal surfers light off (black) and light on (red). The trajectories are started at position (0, 0).  The self-propulsion is isotropic and the dynamics is a persistent random-walk. The particles self-propel at a given velocity $V$ which direction is randomized by thermal fluctuations, over a time scale $\tau_r$. {\bf D.} Mean Square Displacement (MSD) $\Delta L^2$  averaged for a dozen particles (red symbols). The MSD is well described by a persistent random-walk dynamics (Eq. \ref{eq-1}, blue dashed line). For short times, $t<<\tau_r$, the trajectory is ballistic and $\Delta L^2 \propto \Delta t^2$. The self-propelled particles exhibit an enhanced effective diffusion at long times $\Delta L^2\propto 4 D_\mathrm{eff} \Delta t$. The extrapolation of the enhanced diffusion regime allows an alternate determination of the persistent length of the motion (Eq. \ref{eq-2}, black dashed line). {\bf E.} Self-propulsion velocity of hematite composite surfers as a function of the light intensity. Blue symbols for activation by blue [430-490nm] light, and full violet  circle for activation by the  UVA-violet [370-410nm] light. The red dashed line is a fit of the experimental data by a Michaelis-Menten kinetics, typical of enzyme catalysis. {\bf F.} Self-propulsion velocity  as a function of the Debye length  of the solution, varied by addition of Sodium Chloride salt (black symbols) and withdrawing of the SDS surfactant in the solution (blue symbol). The experimental results agree with $V\propto \lambda^2_D$ (red dashed line), expected for phoretic interfacial transport, stressing the role of electrostatics in the system.}
\label{fig4}
\end{figure}
\vspace*{-10pt}


\subsubsection{Individual Dynamics}
  
The dynamics of individual active colloids is investigated measuring the two-dimensional (x,y) motion of  the colloids with a camera (Lumenera Infinity X32 or Edmund Optics-1312M) at a frame-rate between 1 and 50Hz. The position and trajectories of the particle are extracted and reconstructed with a single particle tracking algorithm using a Matlab routine adapted from  Crocker and Grier \cite{Crocker:1996wp}. We use a circular Hough transform and circular shape recognition to determine the position of the center of the particles, and avoid the inaccurate determination of the center otherwise induced by the  contrast of the composite particles containing black and white components.   \\
The trajectories  are then extracted. The mean square displacement (MSD) of the colloids is obtained as $\Delta L^2(\Delta t)=\langle{(\vec{R}(t+\Delta t)-\vec{R}(t))^2}\rangle$ where $\vec{R}(t)$ is the (2D) instantaneous colloid position and the average is performed over time for each individual trajectory and then over an ensemble of trajectories (typically
15).
For  the  activated colloids, the mean square displacement differs drastically from the equilibrium diffusive dynamics. The colloid exhibits ballistic motion at short times,
$\Delta L^2(t)\sim V^2 \Delta t^2$, while at longer times a diffusive regime, $\Delta L^2(t)\sim 4D_{\rm eff} \Delta t$, is recovered with an effective diffusion coefficient $D_{\rm eff}$ much larger than the equilibrium coefficient $D_0$. 
As discussed in \cite{Howse:2007ed ,Palacci:2010hk}, the active colloids are expected to perform a persistent random walk, due to a competition between ballistic motion under the locomotive power (with a constant swimming velocity $V$), and angular randomization of the direction over a persistence time $\tau_r$. In the experiment, we measure a persistent time $\tau_r$ consistent with the thermal Brownian rotational diffusion of the particles at a given radius $R$. The transition between the two regimes occurs at the rotational  diffusion time $\tau_r$ of the colloids. 
The characteristic ballistic length scale is accordingly $L_p=V\times \tau_r$. 
For time scales long compared to $\tau_r$, the active colloids therefore perform a random random walk with an effective diffusion
  $D_{\rm eff}=D_0 +V^2 \tau_r/4$. 
The full expression of the mean squared displacement at any time for a purely 2D motion is obtained as  \cite{Howse:2007ed,Palacci:2010hk} :
\begin{equation}
\Delta L^2(\Delta t)=4D_0\Delta t +\frac{V^2\tau^2_r}{2}[\frac{2\Delta t}{\tau_r} +e^{\frac{-2\Delta t}{\tau_r}}-1]
\label{eq-1}
\end{equation}
 We can accurately fit the experimental data (Fig. 4D, red symbols) with the persistant random walk dynamics [Eq. \ref{eq-1}] (Fig.4D, blue dashed line. The measured persistance time  are in line with the equilibrium Stokes Einstein rotational diffusion. We measure, for example,  $\tau_r=8.0\pm1.5s$ for composite particles with a radius R=$1\mu$m. Note that the  Stokes-Einstein rotational time exhibits a cubic dependence on the particle size $\tau_r\propto R^3$, making it sensitive to the size of the polymer sphere embedding the photoactive material. This can be finely tuned through the synthesis during the growth step and provides an additional source of control in the experiment.\\
The persistence length $L_p$ of  the motion can be extracted from the MSD [Eq. \ref{eq-1}], as the extrapolation of the enhanced diffusion regime at t=0 [Fig.4D, black dashed line]. For $t >> \tau_r$, Eq. \ref{eq-1} rewrites:  
 \begin{equation}
 \Delta L^2(\Delta t>> \tau_r)\sim4D_{eff}\Delta t -\frac{L^2_p}{2}
\label{eq-2}
\end{equation}
  where $L_p=V\tau_r$ is the persistence length.\\
 The propulsion can be tuned with external parameters: reducing the intensity of the activation light reduces the velocity [Fig.4E], following a Michaelis-Menten kinetics typical of enzyme catalysis \cite{Michaelis:1913um} . Shining the violet light (E$_2$) on composite hematite particles has no effect on the propulsion velocity of the particles in line with the assumption of diffusion-limited regime describing the experiment (see Eq.\ref{Eq:diff2} and Fig.2D) \\
 Finally, the propulsion is altered by the ionic strength: the addition of salt reduces the velocity. Osmotic flows generically exhibit a velocity dependence $V\propto L^2$, due to a balance between a driving force $\propto L$ acting over the interfacial layer of thickness $L$  and the viscous drag $\propto 1/L$ in this layer. The scaling is expected to be modified in the regions where the interfacial layer thickness is comparable with the roughness of the surface, as observed for example for electroosmotic flows \cite{Messinger:2010ht}. Our experimental results are in good agreement with a velocity $V\propto \lambda^2_D$, $\lambda_D$ being the Debye length, defining the range of the electrostatic screening around a charged colloid [Fig.4F]. This stresses the role of electrostatics in the system thus discarding neutral diffusio-phoresis/osmosis as the main transport mechanism in our system. The importance of electrostatic contributions in the propulsion mechanism of platinum-coated Janus particles has been recently pointed and thoroughly discussed  in \cite{Brown:2013tp,Anonymous:b6jSMxtT}. 
 
\section{Emergence of collective effects and Living Crystals} 
\subsection{Living Crystals}
For dilute sample of composite particles, we observe a "gas phase": the particles self-propel with a persistent random walk  and collide each other sometime [movie 6]. The collisions are isotropic, and the behavior of the particles shows no significant difference with the individual  behavior.  Increasing the surface density $\Phi_s$ of the particles in the experiment, we observe the emergence of  "living crystals" [movie 7]. This state is qualitatively very different from the gas phase, particles spontaneously assemble in  mobile crystallines structure, mobile, which exchange particles, collide, break appart and reform. This phase is observed for surface fraction, as low as $\Phi_s>7-10\%$. Following a collision, two crystals rearrange, "heal" and suppress  the existence of a grain boundary [movie 7]. \\
In order to see if our understanding of the main components of the system accounts for these observations, we developed a simple code in matlab of  (i) self-propelled hard disks with a persistent random walk dynamics with (ii) a phoretic attraction between {\it pairs} of disks as observed in the experiment. 

\subsection{Numerical simulations and Algorithm}
\subsubsection{Dynamics} 
We consider a numerical model in which the self-propelled colloids are represented by hard disks propelled with a constant velocity $V_0$ along a direction, diffusing on a circle over a time scale $\tau_r$ governed by rotational Brownian diffusion. It is implemented in the simulations as  a random gaussian noise to the propulsion angle. The variance of the gaussian noise controls the the persistence time $\tau_r$ of the motion [Fig.5A].
\begin{figure}[htbp]
\centering\includegraphics[width=4.2in]{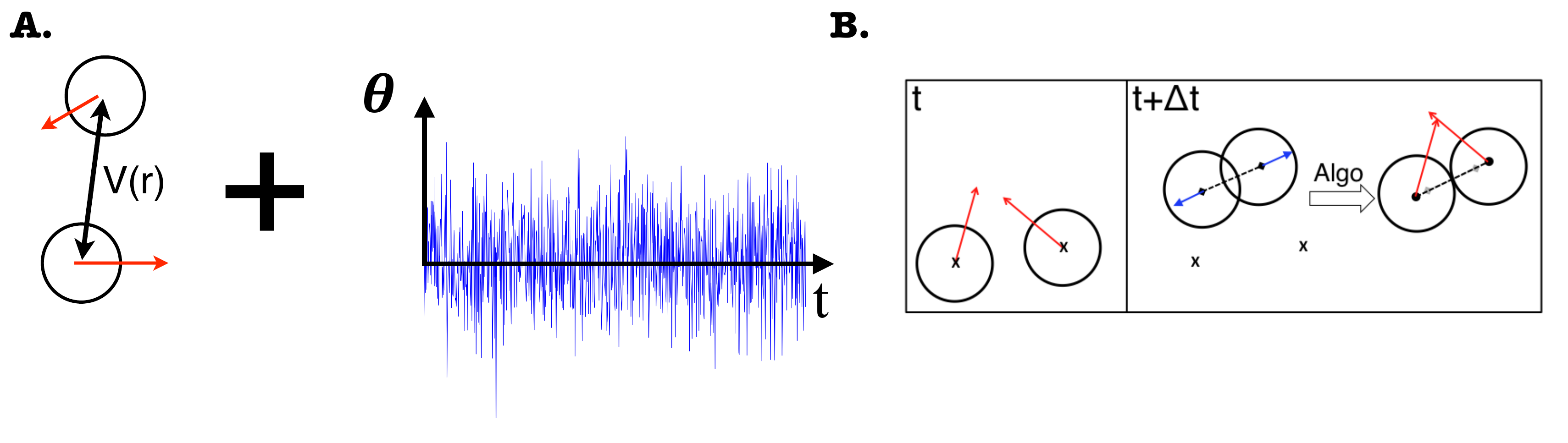}
\caption{{\bf Numerical Simulations.} {\bf A.} We consider 2D simulations of hard disks propelled at a constant velocity $V_0$ (red arrows) with a fluctuating direction of propulsion and an attraction interaction $V(r)$ (black arrow). The noise on the direction is a a Gaussian noise, which amplitude is set by thermal rotational diffusion. {\bf B.} Event-driven algorithm for the collisions. If two particles overlap after a displacement, they are separated by moving each one of them half of the overlap distance along their center-to-center axis. This algorithm does not introduce correlations in the propulsion direction nor collisions and motion.    }
\label{fig5}
\end{figure}

\vspace*{-10pt}
\subsubsection{Interaction between particles}
 We model the phoretic attraction between the particles as a pairwise attractive interaction. For each time step $\Delta t$, the particle i
undergoes a displacement $\Delta R_i$     resulting from its own self-propulsion and the attraction
by the neighbors, for  $j \neq i$:$\Delta R_i = V_0 \Delta t + \Sigma_{j \neq i} V_{att} (r_{i,j} )\Delta t$, with $r_{i,j}$ being the distance between 
particle i and j. The pairwise attraction follows the phoretic attraction  $V_{att} (r_{j,i}) \propto 1 / r^2 _{i,j}$, measured experimentally.  The hard-sphere repulsion between particles is event driven: if a displacement makes two particles overlap, they are separated by moving each one of them half of the overlap distance along their center-to-center axis [Fig.5B]. The displacement perpendicular to the line of centers is preserved. This is an appropriate procedure for low Reynolds number propulsion. It shares forces and hence velocities along the line of centers and preserves tangential velocities. It does not correlate steric effects and motion. Rather it is similar to introducing a sharp repulsive potential between particle surfaces in the very low Reynolds number limit.\\
{\it Limits of this model.} The model assumes pairwise interactions between the particles. The actual attraction should result from the concentration field induced by the resolution of the coupled  diffusion-reaction equation around each particles.  Solving the diffusion-reaction equation of a collection of self-propelled sinks of hydrogen peroxide is a complex problem beyond the scope of this paper. Moreover, the interaction here considered results in an effective potential $E(r)\propto 1/r$ analogous to an unscreened gravitional interaction. The bigger the cluster, the more attractive, eventually leading to a "gravitational collapse". We observe such phenomena, for which the simulations break down and we therefore imposed in the simulations a threshold for the interaction range of 3 particles diameters. This is in line with experimental results where particles do not seem to interact above this distance, but remains questionable in the frame of our model. Our physical picture for the cut-off is that the concentration profile inside the crystal is uniform and flat and that only the particles at the edge of the crystal contributes to the attraction. This phenomenon is not taken into account summing pairwise interactions.
 \subsubsection{Numerical Parameters}
 The simulations are made dimensionless setting the diameter  $\tilde D=1$ and the  the velocity $\tilde V_0=1$ of the self-propelled disks   to unity. This defines a spatial and temporal time step for the simulations.s 
 The parameters in the simulation are fixed in the following range, accordingly to the experimental value:
\begin{itemize}
\item Diameter of the particles:  $\tilde D=1$ 
\item Rotational diffusion time: $\tilde{\tau}_r\equiv \tau_r/\tau=8$ to $50$.
\item Pairwise attraction: $\tilde{V}_{att}(\tilde{r})\equiv -\tilde A/\tilde{r}^2$. 
\end{itemize}
The simulations run with $N=1-1000$ particles in a box of typical size $\tilde L=60$ with periodic boundary conditions with various surface fractions $\Phi_s$ of active particles. The time step for updating  particle positions is $\Delta \tilde{t}=1/200$, and particle-particle pairwise attractions are cut off for interparticle distances greater than  3. 

\subsection{Results and Discussion}
At low density of particles, the simulations show self-propelled disks with a persistent random-walk dynamics. Increasing the surface density of the particle, they form crystals and reproduce qualitatively well the experiment [movie 8]. Following recent  experimental \cite{Narayan:2007bg}  and theoretical \cite{Fily:1x4TMNFH} works, we measured the number fluctuations in the system and showed a transition from normal to giant fluctuations for a critical surface coverage of particles $\Phi^C_S\sim 7\%$ \cite{Palacci:2013eu}. 
This quantitative agreement between the experimental and the numerical results stresses that a simple model of self-propelled particles with a persistence length and attraction between the particles capture the essential ingredients at stake in the experiment. \\
However, one can still envision two {\it scenarii} for the emergence of dense structures in this system. First, it could arise from the light-activated phoretic attraction between the colloids.  If the attraction overcomes the loss of entropy in the dense phase, an equilibrium system spontaneously phases separates into a dense liquid and a solid. It is therefore expected that suspensions of attractive colloids form clusters at equilibrium.\\
 Furthermore, the situation of self-propelled particles coupled by chemical attraction has been studied extensively in biology for bacteria with chemical  {\it chemotactic} sensing since the pioneering work by Keller and Segel \cite{Keller:1970vv}. The  so-called Keller-Segel equation (KS) is a mean field description taking into account the diffusion of bacteria, a drift induced by chemical sensing, and the production and diffusion of a chemoattractant.  An interesting feature of the KS description is that it exhibits singular solutions and a "chemotactic collapse" of the structure into a single or many dense aggregates, above a threshold, the {\it Chandrasekhar} number $N_C$ \cite{Brenner:1998jc}. This description was used  to discuss the emergence of clustering in a collection of Janus particles coupled by diffusio-phoretic chemical attraction and performed by others  \cite{Theurkauff:2012jo}. \\
Alternatively, it has been  pointed out that self-propelled particles lacking an alignment rule exhibit collective behavior and form dense dynamical clusters in equilibrium with a gas phase. This behavior arises from a self-trapping mechanism: self-propelled particles with a persistent time and colliding head on, arrest each other due to the persistence of their orientation [Fig.6A]. Increasing the surface fraction of particles, this simple mechanism leads to a dynamic phase transition from a gas phase  of hot colloids \cite{Palacci:2010hk} to a  dense state, resulting from the "traffic jam" of the persistent self-propelled particles \cite{Fily:1x4TMNFH, Redner:2013jo,Berthier:2013bf,Bialke:2012cw,Menzel:2013gs,Mognetti:2013jz, Stenhammar:2013kb,Bialke:2013gw,Fily:2014jp}. The emergence of arrested phase due to density-dependent mobility has been discussed theoretically in the context of bacteria by Tailleur and Cates \cite{Tailleur:2008kd}. The role of the self-trapping mechanism for the emergence of clustering was shown in the recent and remarkable experiments by Buttinoni {\it et al} \cite{Buttinoni:2013de} where they used "large" $4\mu m$ self-propelled carbon-coated Janus colloids, which self-propel under illumination in a near-critical water-lutidine mixture \cite{Buttinoni:2012ho}, and for which the caps can be optically resolved, indicating the direction of self-propulsion. They showed that the particles in a clusters are arrested, heads-on.  \\

\begin{figure}[htbp]
\centering\includegraphics[width=4in]{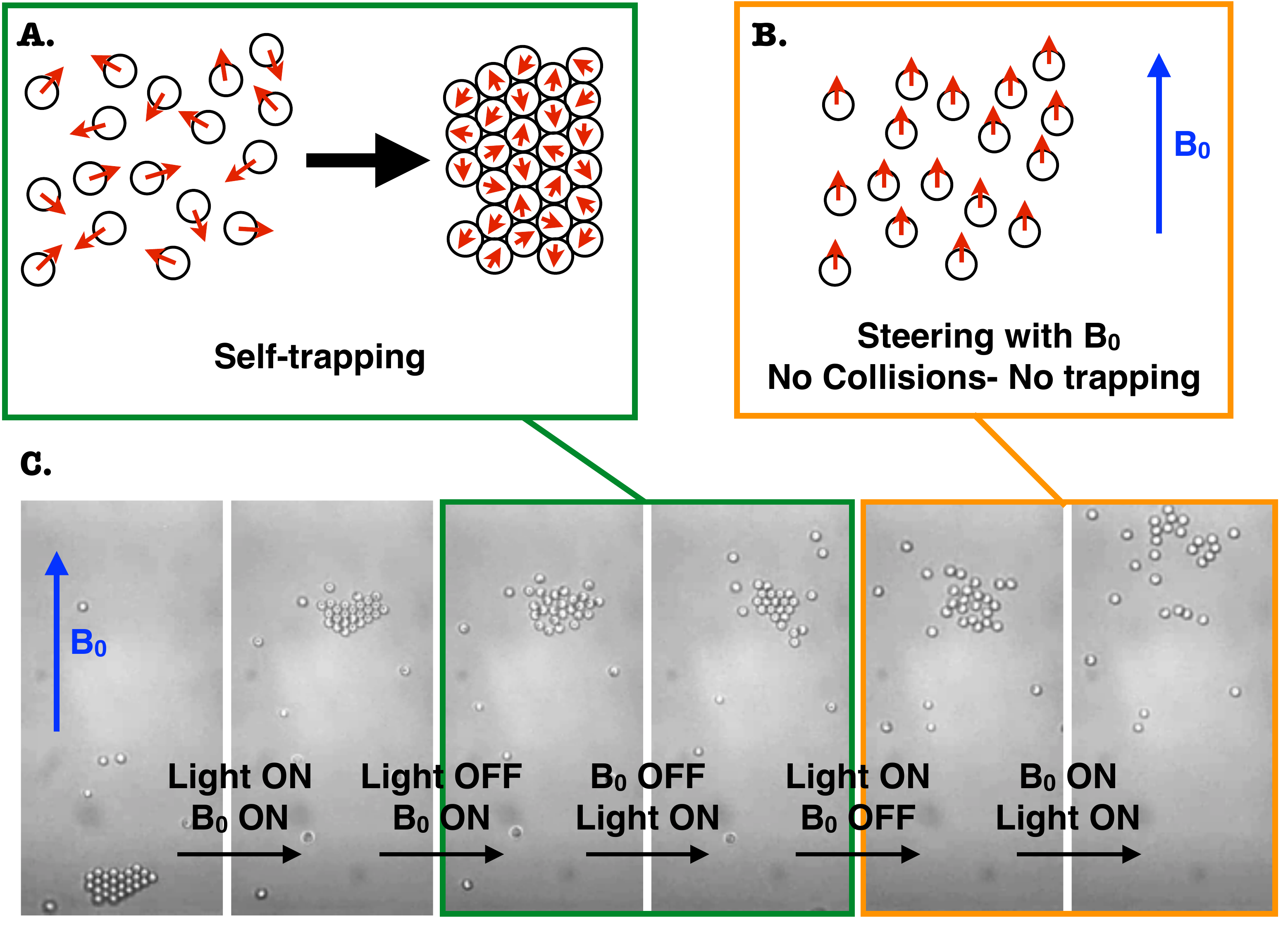}
\caption{{A. Self-trapping Mechanism} Self-propelled particles with a persistent length spontaneously exhibit a dynamic phase transition and form dense clusters. They collide head on, arrest each other due to the persistence of their orientation leading to self-trapping. {\bf B.} Using an external magnetic field to direct the hematite colloidal surfers, the particles all go in the same direction, they do not collide, self-trapping is suppressed.  {\bf C.} In the experiment, the light activation induces attraction and persistent self-propulsion, thus entangling the role of attraction and self-trapping in the formation of the crystals. A magnetic field is used to suppress the collisions and shows the importance of the collisions in the system to form the crystals. A crystal is first formed and steered using a uniform magnetic field $B_0\sim 1$mT. The light is turned off, the magnetic field being on, the system is at equilibrium and the crystal melts. The magnetic field is turned off, and the light is activated. The particles self-propel, collide and reform the crystal. The light is turned off as well as the magnetic field, the crystal melts. Now, the magnetic field is on before the light is activated. The particles self-propel in the same direction, they do not collide, the crystal does not reform. This shows that the collisions are important to observe in the timescale of the experiment, the emergence of the crystals in the system. }
\label{fig6}
\end{figure}

\vspace*{-10pt}

In our experiments, however, the light activation induces both  (i) an attraction as well as (ii) the persistent self-propulsion entangling the importance of an equilibrium-like crystallization with a non equilibrium dynamic phase transition. In order to disentangle the two contributions, we take advantage of the the magnetic properties of the hematite to steer the particles. Forcing the particles to all go in the same direction, we suppress the collisions between the particles thus testing  the importance of  the self-trapping  in the formation of our living crystals [Fig.6B]. \\
We start with a crystal of self-propelled particles formed along a wall, since collisions are favorable along the 1D system defined by the wall. We   steer it  away from the wall using a uniform magnetic field $B_0\sim 1\mathrm{mT}$. Turning off the light as well as the magnetic field, the crystal melts, the particles diffusing away. Turning on the light, the magnetic field remaining off, the individual particles self-propel in every direction, {\it collide} and reform the crystal.  The light is turned off, and the  crystal melts again.  Now the magnetic field $B_0$ is actuated before the light is turned on again. The particles  all propel all along the same line, the direction of $B_0$,  they do not collide and do not form the crystals. This shows that within the time scale of the experiment, the phoretic attraction is not sufficient to account for the formation of the living crystals [Fig.6C]. \\
To summarize, the collisions and self-trapping make the system phase-separate and form aggregates. The presence of the attraction orders these aggregates into 2D crystals and shifts the threshold for the dynamical transition to a low surface density of particles $\Phi_s\sim 7-10\%$, in line with the numerical results by Redner {\it et al} \cite{Redner:2013be}. In the absence of attraction between particles, the dense state develops for larger surface fractions,  typically $\sim 30-40\%$ for self-propelled hard-disks with no alignment as discussed theoretically by Fily and Marchetti  a  \cite{Fily:1x4TMNFH}, and confirmed experimentally and numerically by \cite{Buttinoni:2013de}.

\subsection{Living Crystals vs Swarming and Flocks behavior}
We do not observe experimentally or numerically the swarming behavior predicted theoretically by many authors \cite{Toner:2005bj, Simha:2002eg, Chate:2006en,Gregoire:2004ic} for nematic self- propelled particles similar in spirit to the Vicsek model \cite{Vicsek:1995fk}: transition to a flocking behavior with coherent groups of particles moving in the same direction, swarming or formation of traveling bands otherwise recently observed experimentally by the group of Bausch \cite{Schaller:2010cq} or more recently the large scale vortices observed by Sumino {\it et al} \cite{Sumino:2012dw} or by the group of Bartolo with self-propelled rollers colloids \cite{Bricard:2014jq}, or active nematics \cite{Sanchez:2013gt}. In all those works, the self propelled particles are polar with nematic interactions, i.e. they align their velocity vectors in direction, as prescribed by "Vicsek's rule". The role of the nematic alignment due to the collisions of rod-like microtubules have recently been pointed by \cite{Sumino:2012dw} to explain the emergence of long range interactions and vortices at high density (or in an granular context by Deseigne et al \cite{Deseigne:2010p1625}). In our experiment, due to the isotropic shape of the particles we do not observe any nematic interaction of the self-propelled particles, thus defining a different   class of systems and non-equilibrium phases than the Vicsek model. \\
To our knowledge, the transition from dynamic clusters emerging from a self-trapping mechanisms, for particles without alignment, to flocking behavior, for active nematic,  has not been studied theoretically. It would be interesting to see how altering the alignment mechanism, e.g. altering the shapes of the self-propelled particles,  can induce a transition from one class of collective behavior to an other.

\section{Conclusion}
In the paper we showed how we could harness the photo properties of hematite and titania semi-conductors to design self-propelled colloids in a solution of hydrogen peroxide fuel. The photocatalytic  decomposition of hydrogen peroxide  is triggered by a light with a wavelength above the bandgap of the material. It generates a chemical gradient around the particles, which induces an osmotic self-propulsion of the particle along the substrate and a phoretic attraction between the particles.  We demonstrate here, for the first time, a wavelength-dependent activation in a mixture of different types of self-propelled particles. The engineering of such  colloids provides a general route for designing new families of self-propelled particles.\\
  We show that a collection of these particles spontaneously assemble in living crystals, mobile which form, heal, break apart and reform, and could reproduce this with numerical simulations of a simple model. Using a magnetic field to direct the particles with a iron oxide, hematite, component, we show that the collisions are central in the observed formation of the crystals. This self-trapping mechanism is a non equilibrium effect and points towards the emergence of novel organization principles in non equilibrium systems. The transition from self-trapping to flocks is a open question, which could be addressed, for example, changing the shape of the self-propelled particles.
\section*{Acknowledgment}
We thank Aleks Donev and his group for the numerical simulations of the phoresis/osmosis induced by  a sink of chemical [Fig.2A]. 
This work was supported by the Materials Research Science and Engineering Centers program of the NSF under award number DMR-0820341 and by the U.S. Army Research Office under grant award no. W911NF-10-1-0518. We acknowledge partial support from the NASA under grant award NNX08AK04G.
GRY and SHK acknowledges support from Korean NRF grant (2010-0029409) and Human Resources Development Program (No.20124010203270) of KETEP. JP and PMC acknowledges support from the Moore foundation. 

\section*{Materials and Methods}
\subsection{Synthesis of the Active Colloids}
\subsubsection{Hematite Cubes}
Hematite ($\alpha$ Fe$_{2}$O$_{3}$) cubic colloids were prepared following the method described by Sugimoto et al. in \cite{Sugimoto:1993bf}. Briefly a ferric hydroxide gel was prepared by mixing 100mL of aqueous NaOH (6M) with 100mL of FeCl$_{3}\times$ 6H$_{2}$O (2M) and aged in a sealed Pyrex bottle at 100$^{\circ}$C. After  8 days the gel changed into a thick reddish sediment which was repeatedly washed in deionized water to reveal the colloidal cubes. From electron microscopy pictures, we measured an average particle size of $600$~nm with a typical polydispersity of 3$\%$.
\subsubsection{Titania Microspheres}
Titania (TiO$_2$) colloids were prepared by hydrolysis and condensation reaction of titanium isopropoxide (TTIP) with dodecylamine (DDA) as a catalyst in a co-solvent of methanol/acetonitrile \cite{Tanaka:2009bt}.
In a typical synthesis, 0.384 ml of water was added to a solution consisting of  103 ml of methanol and 32 ml of acetonitrile. Then 0.7 g of DDA was dissolved in the mixture, followed by 1.07 ml of TTIP.  The mixture  was left under stirring for 12 h. The resulting  titania suspension was centrifuged at 1500 rpm for 10 min and the sediments were  washed  three times with  methanol. The the partiles were finally dried and  calcinated  at 500C for 5h.

\subsubsection{Encapsulation of Hematite or Titania}
To embed the hematite or titania component into larger spherical particles we added  $25~\mu$L of NH$_{3}$ 28$\%$ to a 30 mL aqueous suspension of hematite particles ($\approx2\%$ wt) followed by 100$\mu$L of 3-methacryloxypropyl trimethoxysilane (TPM, $\ge98\%$ from Sigma-Aldrich). The reaction mixture was kept under vigorous stirring and sampled every 15 minutes to monitor the particles' growth. The reactor is fed with more TPM (100$\mu$L of TPM for each addition) at intervals of approximately 1~h until the particles reached the desired size. Finally 0.5~mg of 2,2'-azo-bis-isobutyrylnitrile (AIBN, Sigma-Aldrich) were added and the mixture heated to 80$^{\circ}$C for 3h  to harden the particles.
After the synthesis the particles were  cleaned and separated from secondary nucleation by sedimentation and were finally resuspended in deionized water. The surface zeta potential in water at a pH of 9 was measured to be -70~mV.


\end{document}